\documentclass[twocolumn,epsfig,APS,floats,showpacs]{revtex4}
\usepackage{graphicx}
\usepackage{color}

\usepackage{fancyhdr}
\fancypagestyle{titlepage}{
\fancyhead[R]{Journal of  Physics: Condensed Matter \textbf{26}, 415501 (2014)}}
\begin{document}

\title{Crystallization Characteristics and Chemical Bonding Properties of Nickel Carbide Thin Film Nanocomposites}

\author{Andrej Furlan$^{1,*}$, Jun Lu$^{1}$, Lars Hultman$^{1}$, Ulf Jansson$^{2}$ and  Martin Magnuson$^{1,**}$}

\affiliation{$^1$Department of Physics, IFM, Thin Film Physics Division, Link\"{o}ping University, 
SE-58183 Link\"{o}ping, Sweden.}

\affiliation{$^2$Department of Chemistry - \AA{}ngstr\"{o}m Laboratory, Uppsala University, Box 538, SE-751 21 Uppsala Sweden.}

\affiliation{$^*$Present address: AMB Industry, Kvarnv\"{a}gen 26, 361 93 Broakulla, Sweden.}

\date{\today}

\begin{abstract}
The crystal structure and chemical bonding of magnetron-sputtering deposited nickel 
carbide Ni$_{1-x}$C$_{x}$ (0.05$\leq$x$\leq$0.62) thin films have 
been investigated by high-resolution X-ray diffraction, transmission electron microscopy, 
X-ray photoelectron spectroscopy, Raman spectroscopy, and soft X-ray absorption 
spectroscopy. By using X-ray as well as electron diffraction, we found carbon-containing 
hcp-Ni (hcp-NiC$_{y}$ phase), instead of the expected rhombohedral-Ni$_{3}$C. 
At low carbon content (4.9 at\%) the thin film consists of hcp-NiC$_{y}$ 
nanocrystallites mixed with a smaller amount of fcc-NiC$_{x}$. The average 
grain size is about 10-20 nm. With the increase of carbon content to 16.3 at\%, 
the film contains single-phase hcp-NiC$_{y}$ nanocrystallites with expanded 
lattice parameters. With further increase of carbon content to 38 at\%, and 62 
at\%, the films transform to X-ray amorphous materials with hcp-NiC$_{y}$ 
and fcc-NiC$_{x }$ nanodomain structures in an amorphous carbon-rich 
matrix. Raman spectra of carbon indicate dominant $sp^{2}$
hybridization, consistent with photoelectron spectra that show a decreasing amount 
of C-Ni phase with increasing carbon content. The Ni $3d$ - C $2p$ 
hybridization in the hexagonal structure gives rise to the salient double-peak 
structure in Ni $2p$ soft X-ray absorption spectra at 16.3 at\% that changes 
with carbon content. We also show that the resistivity is not only governed by 
the amount of carbon, but increases by more than a factor of two when the samples 
transform from crystalline to amorphous. 

\end{abstract}

\maketitle

\section{introduction}

Transition metal carbides are useful in various applications ranging from wear 
and oxidation resistant protective coatings to low friction solid lubricants \cite{1,2}. 
This flexibility is due to the nanocomposite nanocrystalline/amorphous-C structure 
that governs the coating's properties depending on the amount of amorphous matrix, 
and crystallite size of the carbide \cite{3,4}. Early transition metals such as Ti, 
Zr, and V form strong covalent metal carbon bonds often in cubic crystals in contrast 
to late transition metals such as Fe and Ni that form less strong Me-C bonds \cite{5}. 
The late transition metals usually form completely amorphous or mainly amorphous 
materials with complex nanocrystallites above a threshold value around 20 at\%. 
An important exception is the Ni-C system, where metastable rhombohedral-Ni$_{3}$C 
nanocrystallites are easily formed in a large composition range \cite{6,7}, and it is 
more difficult to form completely amorphous films [8]. Using RF sputtering \cite{6}, 
partly amorphous Ni$_{1-x}$C$_{x}$ films with Ni$_{3}$C 
crystallites embedded in an amorphous Ni$_{1-x}$C$_{x}$ phase 
have previously been obtained for \textit{x}=0.35 \cite{6}. Amorphous Ni-C films have 
also been obtained for \textit{x}\texttt{>}0.5 using reactive co-sputtering with 
CH$_{4}$ as a carbon source \cite{7}. 

Metallic Ni usually has a cubic fcc structure with space group Fm-3m. 
In addition, the hcp-Ni phase with space group P63/mmc has been reported \cite{9,10}. Pure hcp-Ni 
metal is very unstable, and most previous investigations lack a material composition 
analysis [6,7]. One study showed that nano-crystallites of hcp-Ni metal could be 
synthesized, and likely stabilized by carbon, but is easily transformed into fcc-Ni 
metal when the size extend more than 5 nm \cite{11}. Because both hcp-Ni with space 
group P63/mmc and Ni$_{3}$C with space group R-3cH contain carbon and 
have very similar crystal structures, most previous work did not experimentally 
show how to discriminate hcp-Ni from Ni$_{3}$C [12,13]. Only recently, 
Schaefer \textit{et al.} \cite{14} were able to distinguish between hcp-Ni and Ni$_{3}$C 
structures by means of low-angle X-ray diffraction. Schaefer pointed out that all 
previously reported hcp-Ni contains carbon, and should be described as rhombohedral 
Ni$_{3}$C \cite{13,14} in a superstructure with interstitially ordered carbon. 
The superstructure can be approximated by a hexagonal subcell that is nearly identical 
in size to that of hcp-Ni with lattice constants \textit{a}=2.682  \AA{}, and \textit{c}=4.306 
 \AA{} \cite{15}.

In this work, we investigate the nanocomposite-to-amorphous structure, and the 
nature of chemical bonding between Ni, and C for a range of C concentrations (0.05$\leq$x$\leq$0.62) 
in magnetron sputtered Ni$_{1-x}$C$_{x}$ films. As a non-equilibrium 
process, magnetron sputtering may increase the solubility of C into Ni, and the 
carbide phase of the film structure may be influenced by the total C content. By 
employing a combination of X-ray diffraction, high-resolution transmission electron 
microscopy (HR-TEM), X-ray photoelectron spectroscopy (XPS), Raman, and soft X-ray 
absorption spectroscopy (XAS), we analyze the Ni carbide, and Ni metal-like nanocrystalline 
to amorphous contributions to the structure, and the dependence on the carbon content. 
In particular, we identify the crystallization of the Ni-C system into hcp-Ni and 
fcc-Ni by combining X-ray diffraction (XRD) with HR-TEM for low carbon contents. 
We show that the samples do not form a superstructure of Ni$_{3}$C with 
ordered carbon as previously thought. XPS characterization gives a quantitative 
analysis of the compositions of different Ni-C phases with particular emphasis 
on the variation of the carbon content in the carbide phase. The electrical resistivity 
of the Ni$_{1-x}$C$_{x}$ films is correlated to the amount 
of C-Ni and C-C bonds, the degree of crystallization, and the total carbon content.

\section{Experimental details}

\subsection{Synthesis and deposition}
All the investigated films were deposited by dual dc magnetron sputtering in ultra 
high vacuum (UHV) on single-crystal Si(001) (10x10 mm) subtrates. Prior to deposition, the substrates were cleaned in ultrasound baths of acetone and isopropyl alcohol. During deposition, the substrates biased to -50 
V, and preheated to 250$^{\circ}$C from the back side by a resistive heater built-into 
the substrate holder. 
This made it possible to synthesize the films with a high 
degree of purity, and with precisely tuned composition. The Ni$_{1-x}$C$_{x}$ 
thin films were deposited in an UHV chamber with a base pressure of 10\textsuperscript{-9} 
Pa from a double current regulated 2 inch magnetron setting in an Ar discharge 
generated at 3.0 mTorr, and with gas flow rate of 30 sccm. The magnetrons were 
directed towards a rotating substrate holder at a distance of 15 cm. As separate 
sputtering sources, graphite, and a non-elemental Ni+C target were used (99.999\% 
pure C, and 99.95\% pure Ni). To enable the magnetic field from the magnetron to 
reach the plasma through the ferromagnetic Ni target, a segmented design was used 
in, which a circular center part of the target was removed, and placed on a graphite 
plate. In this way, simultaneous sputtering of Ni, and C from the same target was 
accomplished \cite{4}. The tuning of the film composition was achieved by keeping the 
graphite target at a constant current of 300 mA, and tuning the current on the 
Ni target. The resulting thicknesses of the as-deposited coatings were 740 nm ((\textit{x}=0.05), 
635 nm ((\textit{x}=0.16), 309 nm ((\textit{x}=0.38), 250 nm ((\textit{x}=0.62) and 200 nm ((\textit{x}=1.0: a-C) as determined by XRR. 

\begin{figure}
\includegraphics[width=80mm]{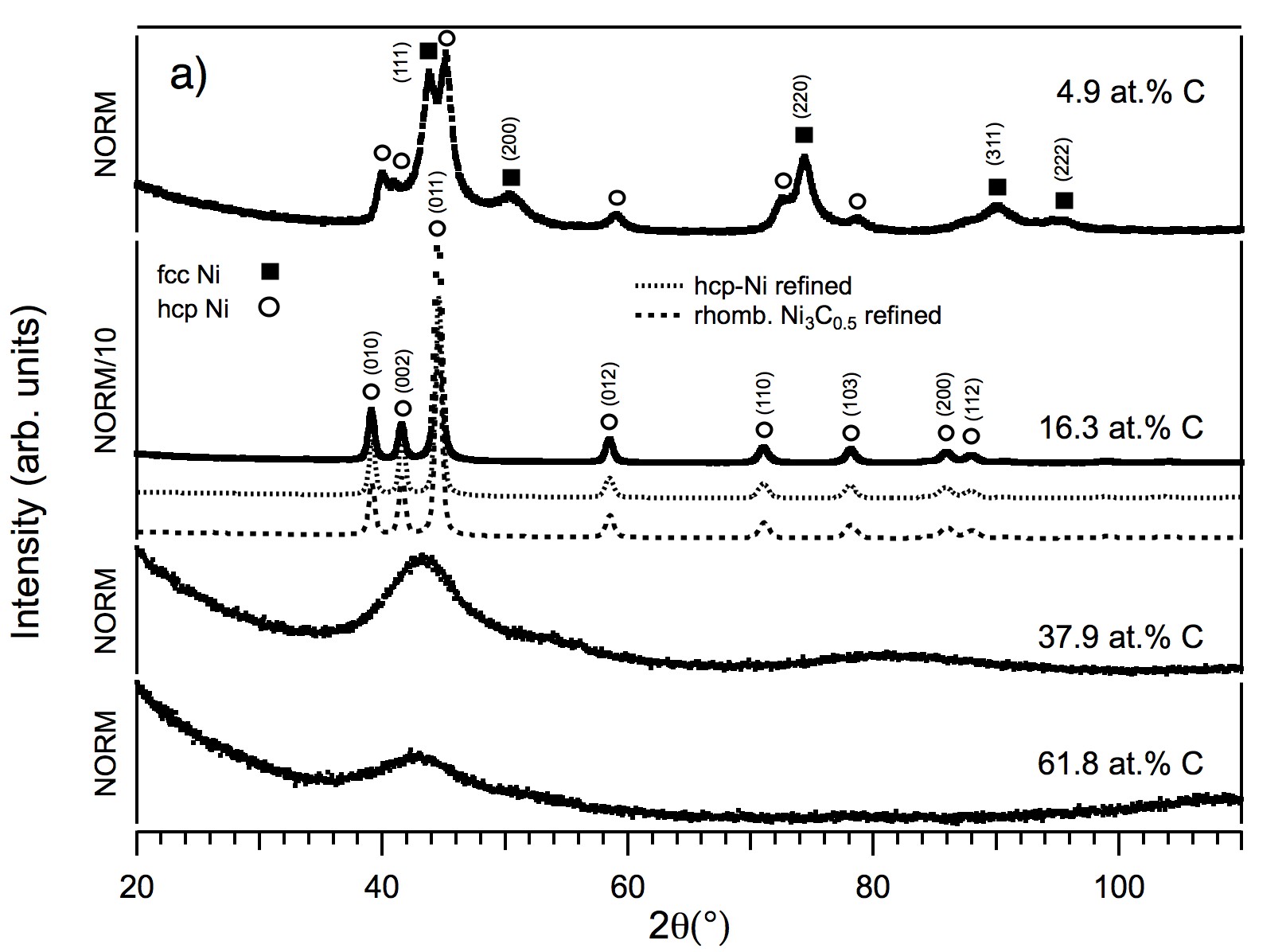} 
\includegraphics[width=80mm]{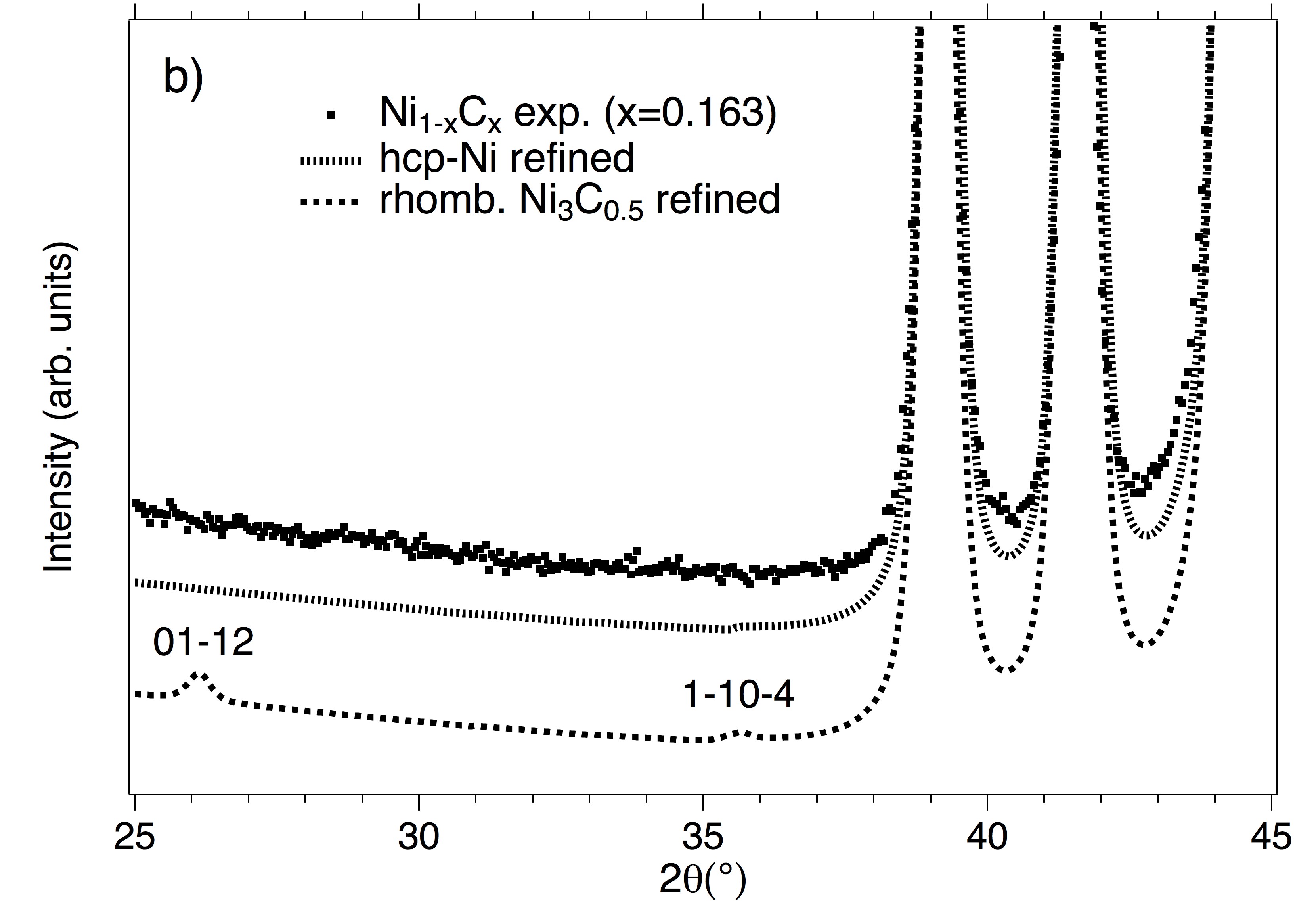} 
\vspace{0.2cm} 
\caption[] {a) X-ray diffractograms of Ni$_{1-x}$C$_{x}$ films with C content ranging from 4.9 at.\% to 61.8 at.\%. b) Enlargement of the 
XRD data of the 16.3\% C sample at low angles in comparison to Rietveld refinement 
of hcp-Ni and rhomb. Ni$_{3}$C$_{0.5}$ with a similar composition as the 16.3 \% sample.}
\label{XRD}
\end{figure}

\subsection{Characterization}
The structural properties of the thin films were determined by high-resolution 
XRD analysis. In order to avoid diffraction signal from the Si substrate, grazing 
incidence (GI) XRD measurements were carried out on a PANanalytical EMPYREAN using 
a Cu \textit{K$\alpha$} radiation source, and a parallel beam geometry with a 2\textsuperscript{o}  
incidence angle to avoid substrate peaks and minimize the influence of texture. 
Each XRD scan was performed with 0.1\textsuperscript{o} resolution, 0.05\textsuperscript{o} 
step length with a total of 1800 points for 6 hours. 

HR-TEM, and selected area electron diffraction (SAED) were performed by using a 
Tecnai G\textsuperscript{2} 20 U-Twin 200 kV FEGTEM microscope. Cross-section samples 
were mechanically polished, and ion milled to electron transparency by a Gatan 
Precision Ion Polishing System (PIPS).

The chemical compositions of the films were determined by X-ray photoelectron spectroscopy 
(XPS) using a Physical Systems Quantum 2000 spectrometer with monochromatic Al 
\textit{K$_\alpha$ }radiation. Depth profiles of the films were acquired by rastered 
Ar\textsuperscript{+}-ion sputter etching over an area of 2x2 mm\textsuperscript{2} 
with ions being accelerated by the potential difference of 4 kV. The high-resolution 
scans of the selected peaks were acquired after 6 min, 30 min, 45 min of Ar\textsuperscript{+}-ion 
sputter etching with ions being accelerated by the potential difference of 4 V, 
500 V, and 200 kV, respectively. The XPS analysis area was set to a diameter of 
200 $\mu$m and the step size to 0.05 eV with a base pressure of 10$^{-9}$ Pa during all measurements. The peak fitting was made by Voigt shape functions to account for the energy resolution of the instrument and chemical disorder (Gaussian part) and the lifetime width of the photoionization process (Lorenzian part).

Raman scattering spectroscopy was used in order to correlate the nanostructuring, 
and $sp^{2}$/$sp^{3}$ ratio of the films to the carbon concentration. The Raman spectra were acquired 
at room temperature in the range 800-1900 cm\textsuperscript{-1} in the back scattering 
configuration using UV 325 nm laser excitation.

X-ray absorption spectroscopy (XAS) measurements were performed in total fluorescence yield (TFY) mode 
at the undulator beamline 
I511-3 at MAX II (MAX-IV Laboratory, Lund, Sweden), comprising a 49-pole undulator, 
and a modified SX-700 plane grating monochromator \cite{16}. The measurements were made 
at a base pressure lower than 6.7*10\textsuperscript{-7} Pa. The XAS spectra were measured at 5\textsuperscript{o} 
grazing incidence angle from the surface plane and a detection angle of 30\textsuperscript{o}  from the incident photon direction. All samples were measured in the same geometry with energy resolutions of 0.2, and 0.1 eV at the Ni $2p$, and C $1s$ absorption edges, respectively. The XAS spectra were normalized to the step before, and after the 
absorption edges and corrected for background and self-absoption effects \cite{16a} 
with the program XANDA \cite{16b} in Fig. 6, and 7.

Cross-sectional scanning electron microscopy (SEM) images were obtained in a LEO 
1550 microscope using accelerating voltages of 15 kV in \textit{in-lens} imaging 
mode. The obtained images were used for thickness measurements, and structural 
analysis of the coatings. 

Sheet resistance measurements were made with a four-point probe ``4-dimensions'' 
280C. For each sample, four readings were made with a different 4-sensor orientation 
around the center of the sample. As a final value of the electrical resistivity, 
a mean value over four measurements was made. Each set of measurements on a sample 
showed similar values indicating negligible influence from surface oxide.

\begin{figure}
\includegraphics[width=87mm]{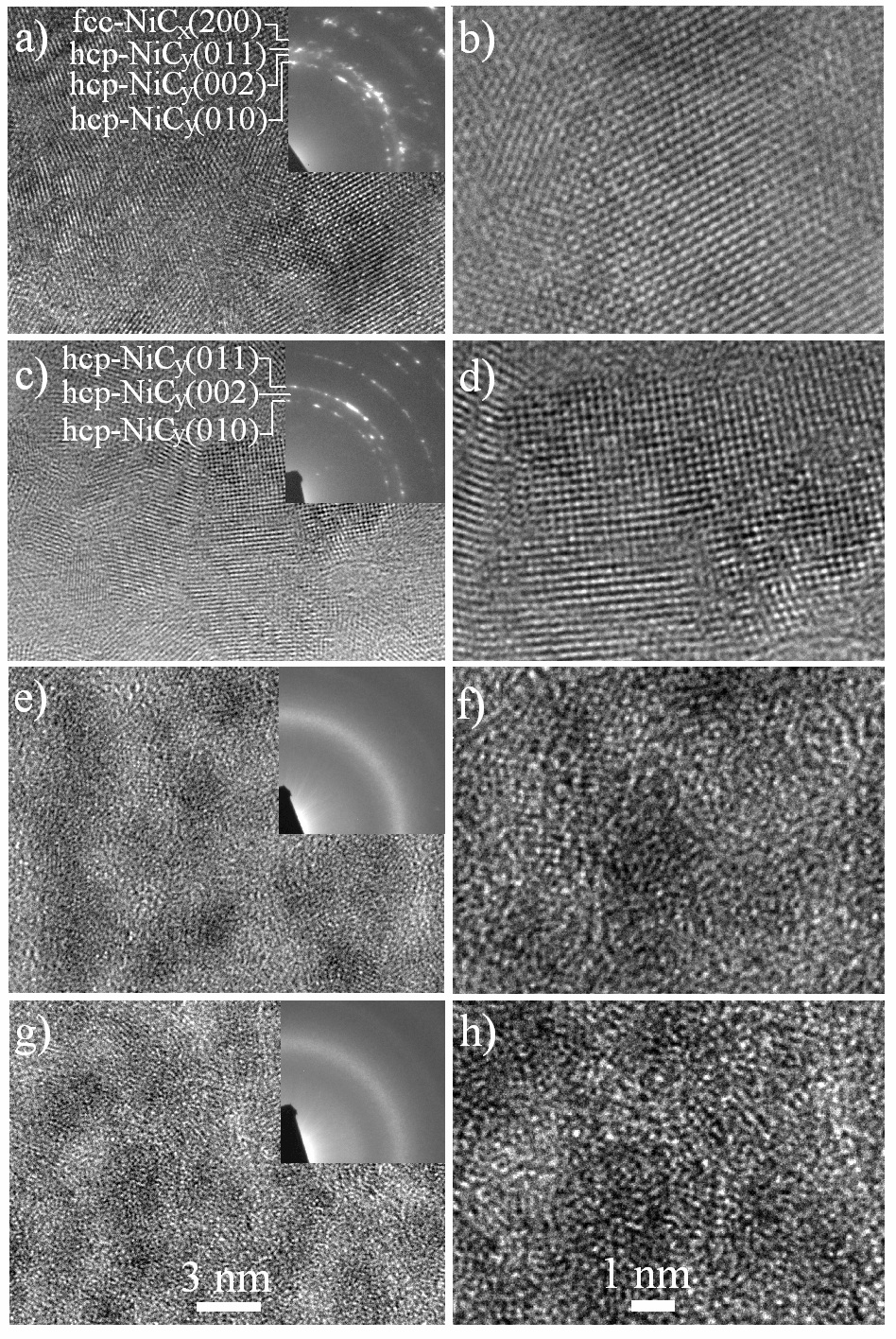} 
\vspace{0.2cm} 
\caption[] {HR-TEM micrographs of low (left), and high (right) resolution 
with corresponding SAED patterns for the Ni$_{1-x}$C$_{x}$ 
films with increasing \textit{x} values, top to bottom; a,b) 0.05, c,d) 0.16, e,f), 
0.38, and g,h) 0.62. The diffraction rings for the fcc-NiC$_{x}$ (200) and hcp-NiC$_{y}$ (011), (002), (010) reflections are indicated in the SAED.}
\label{HR-TEM}
\end{figure}

\section{Results}

\subsection{X-ray diffraction (XRD) and high-resolution transmission electron microscopy (HR-TEM)}
Figure 1a shows X-ray diffractograms (XRD) performed to characterize the microstructure 
of the nickel carbide Ni$_{1-x}$C$_{x}$ films for \textit{x}=0.05, 
0.16, 0.38, and 0.62. The XRD data were refined by the Rietveld method using the 
MAUD program \cite{18}. Five peaks in the top diffractogram (\textit{x}=0.05) are indexed 
as fcc-NiC$_{x}$ structure (space group Fm-3m) with a lattice parameter 
of \textit{a}=3.610(1) \AA{}. This lattice parameter is larger than that of fcc-Ni 
metal (3.524 \AA{}) \cite{17}, but smaller than that of fcc-NiC (4.077 \AA{}) \cite{18}. Interpolation 
of these values yield an estimated phase composition of fcc-NiC$_{x}$, 
where \textit{x}=0.23-0.30. However, the lattice parameter also depends on the 
nano-structured grain size of fcc-NiC$_{x}$ that has larger cell parameters 
than in the case of bulk materials. Therefore, \textit{x}$\leq$0.30 for fcc-NiC$_{x}$ 
in the 0.05 sample, represents an upper limit of the composition.

The other marked reflections in the top diffractogram can be indexed by either hcp-Ni or rhombohedral Ni$_{3}$C structures. From a structure point of view, rhombohedral Ni$_{3}$C has the same Ni position as hcp-Ni but the ordered interstitial C atoms create additional reflections 01-12 and 1-10-4, which are absent for the hcp-Ni structure. Thus, the rhombohedral structure can be excluded based on the absence of 01-12 and 1-10-4 reflections \cite{19,20,21,21b}. Furthermore, compared to the pure hcp-Ni phase, the slight peak shift to low angle indicates an expansion of the lattice, due to the carbon occupation. With increasing carbon content, the cell parameters are further increased (see the second diffractogram of the 16.3 at \% C sample in Fig. 1(a)). Thus, the phase with hcp-Ni structure should be termed as hcp-NiC$_{y}$ (y \texttt{<} 1) instead of hcp-Ni.

The diffractograms are further analyzed with Rietveld refinement. Both hcp-NiC$_{y}$ and rhombohedral Ni$_{3}$C structures were used in the refinement. For comparison, the same carbon content was used for the two structure models, i.e. hcp-NiC$_{0.67}$ and rhombohedral Ni$_{3}$C$_{0.5}$. The refined results in Fig. 1(b) shows that the 01-12 and 1-10-4 peaks are present in rhombohedral Ni$_{3}$C$_{0.5}$ but absent in hcp-NiC$_{0.67}$. The simulated diffractogram of the hcp-NiC$_{0.67}$ agrees well with our experimental data (see Fig. 1(b)). Moreover, the refinement give rise the accurate lattice parameters of hcp-NiC$_{y}$ as $a$=2.611(2) \AA{} and $c$=4.328(7) \AA{} for the 4.9 at\% C sample and $a$=2.653(7) \AA{}, $c$=4.337(2) \AA{} for the 16.3 at\% C sample, respectively. Fig. 1(a) also shows that with increase of carbon content from 4.9 at\% to 16.3 at\%, the fcc-NiC$_{x}$ disappears and single phase of hcp-NiC$_{y}$ forms. Applying Scherrer's equation for the 16.3 at\% C sample yield a grain size of 23 nm for all diffraction peaks in agreement with the average grain size in the TEM images.
Further increase of carbon content to 37.9 at\% and above leads to amorphous-like structures as shown in Fig. 1(a).

The structural evolution of the films with composition is also observed by HRTEM and SAED as shown in Fig. 2. The sharp SAED in Fig. 2c clearly has absence of 012 and 104 reflections, which confirms the hcp-NiC$_{y}$ structure rather than rhombohedral Ni$_{3}$C structure.
No significant texture is observed in the SAED. The films with low carbon contents 
($x$=0.05, and 0.16) are polycrystalline, and the sharp dots of reflections in the 
corresponding SAED pattern show that the film with 16.3 at\% C consists of hcp-NiC$_{y}$, 
while the film with 4.9 at\% C contains two phases: hcp-NiC$_{y}$, and 
cubic fcc-NiC$_{x}$ (the 200 reflection is consistent with XRD). In contrast, the films with higher carbon contents (38, 
and 62 at\% C) consist of Ni-rich nanocrystalline domains surrounded by amorphous 
carbon-rich matrix domains. The average grain size of the nanocrystalline domains 
is approximately 3-5 nm ($x$=0.38, and 0.62). Although it is difficult to identify 
the exact phases for these X-ray amorphous samples with high carbon content, the 
XRD peak intensity distribution profiles indicate that the strongest broad structure 
at 2$\theta$=42.6$^{\circ}$ includes both hcp-NiC$_{y}$ 010, 002 and 011 reflections, and a fcc-NiC$_{x}$ 111 reflection, respectively. The broad 
structure at 2$\theta$=82$^{\circ}$ is formed by fcc-NiC$_{x}$ 200, and hcp-NiC$_{y}$ 
110 and 103 reflections. Thus, as observed by the HR-TEM in Fig. 2, the samples with 
high carbon contents of 38, and 62 at\% likely consist of both cubic fcc-NiC$_{x}$, 
and hcp-NiC$_{y}$ nanocrystallites both with a small grain size approximately 
3-5 nm.

\begin{figure}
\includegraphics[width=85mm]{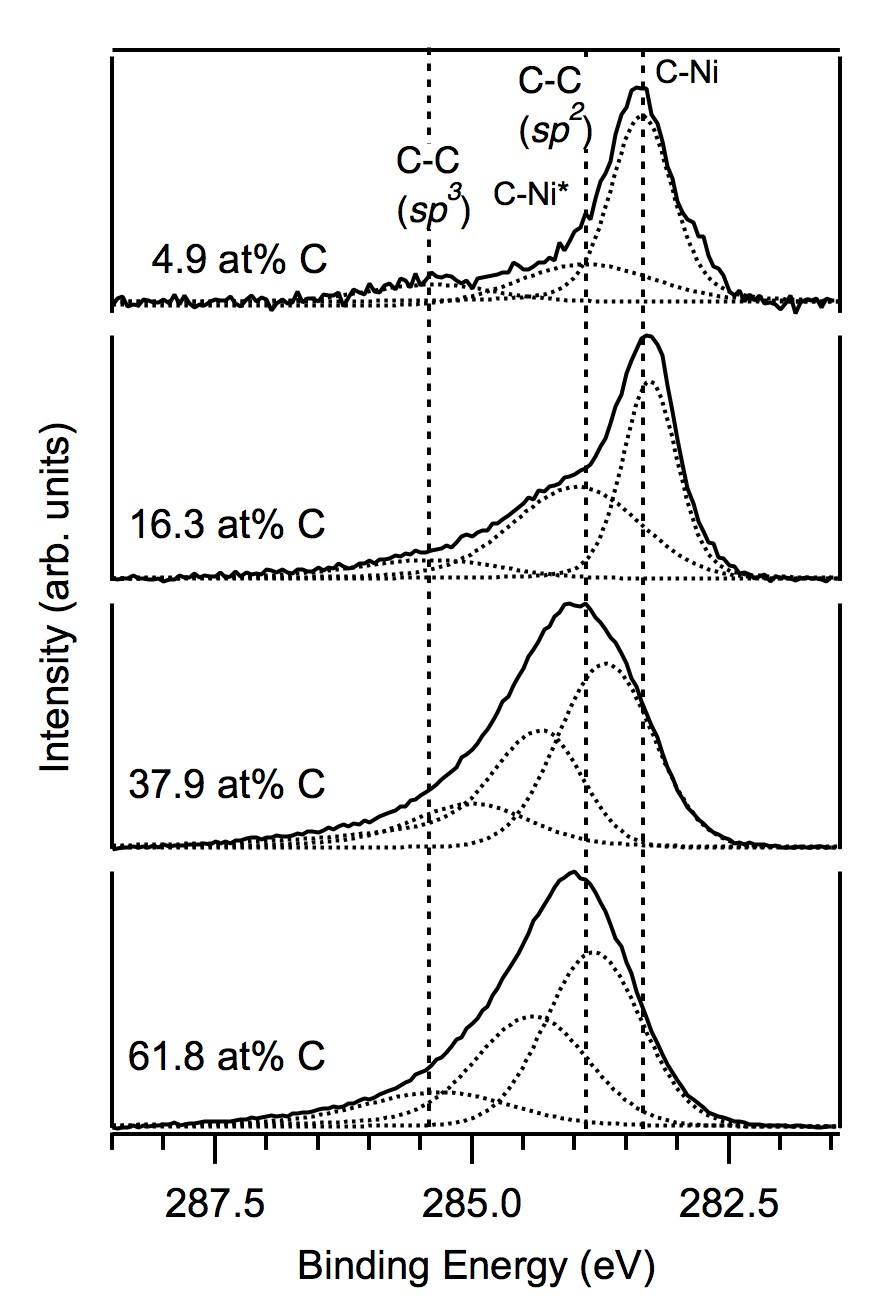} 
\vspace{0.2cm} 
\caption[] {C \textit{1s} XPS spectra of the Ni$_{1-x}$C$_{x}$ 
films with carbon content ranging from 5 at.\% to 62 at.\%. The deconvoluted peaks 
at 283.3, and 285.3 eV, indicated by the dashed vertical lines corresponds to carbidic 
NiC carbon in Ni-C bonds, and free carbon in C-C bonds, respectively. A third structure 
is identified at 283.9 eV, and can be associated with charge-transfer C-Ni* bonds 
or C-C in \textit{sp}\textsuperscript{\textit{2}} hybridized bonds.}
\label{XPS1}
\end{figure}

\subsection{X-ray photoemission spectroscopy (XPS)}
Figure 3 shows C \textit{1s} core-level XPS spectra of the four Ni$_{1-x}$C$_{x}$ 
films with x=0.05, 0.16, 0.38, and 0.62. As observed, at least three peaks are 
required to deconvolute the spectra. A peak at 283.3 eV can be assigned to Ni-C 
bonds, while a second peak at 285.3 eV can be assigned to \textit{sp}\textsuperscript{\textit{3}} 
hybridized carbon (C-C-$sp^{3}$) \cite{22,23}. Between 
these two peaks, a third feature is clearly present. The intensity of the middle 
peak increases with carbon content and is also shifted from about 283.9 eV for 
the most Ni-rich film to about 284.5 eV in the most C-rich film. Most likely, several 
types of carbon is contributing to this feature. Firstly, $sp^{2}$-hybridized 
carbon is known to exhibit a peak at about 0.9 eV lower binding energy than $sp^{3}$-hybridized carbon, i.e. at about 284.0 eV \cite{22}. It is well known that binary sputter-deposited 
metal carbide films often are nanocomposites with a carbide phase in an amorphous 
carbon (a-C) matrix. Previous C $1s$ XPS studies on the a-C phase have shown 
a mixture of $sp^{2}$ - and $sp^{3}$ -hybridized 
carbon. Consequently, it is most likely that a part of the intensity of the feature 
at 283.9-284.5 eV originates from free carbon in an a-C matrix. Secondly, studies 
on sputter-deposited Me-C films have also shown an additional Me-C feature at a 
slightly higher binding energy \cite{3} that can be due to sputter damage of the metal 
carbide grains. Thirdly, a contribution originates from surface Me atoms in the 
carbide grains. This is caused by charge-transfer effects where charge is transferred 
from the metal surface atoms to the more electronegative carbon atoms in the a-C 
matrix \cite{7}. In nanocomposites with very small grains or domains, the relative amount 
of surface atoms is large and will show up as a high-energy shoulder on the main 
C \textit{1s} Me-C peak (denoted Me-C*) \cite{9}. However, for the Ni$_{1-x}$C$_{x}$ 
films, it is impossible to deconvolute the feature at 283.9-284.5 eV into separate 
C-C ($sp^{2}$) and Ni-C* peaks. However, a comparison 
with Ti-C, Cr-C and Fe-C films show that the Me-C* contribution is small in XPS 
compared to the C-C ($sp^{2}$) peak \cite{24,25}. For 
this reason, we assign the entire peak at 283.9-284.5 eV to C in a-C, although 
it will give a slight overestimation of the amount of the C-C ($sp^{2}$) 
phase compared to the NiC carbide phases. The XPS data supports the TEM and XRD 
studies and confirms that the films consist of at least two phases; fcc-NiC$_{x}$ 
and hcp-NiC$_{y}$ carbide nanocrystallites dispersed in an amorphous 
carbon (a-C) phase. The intense C-Ni peak for \textit{x}=0.05 and 0.16 is an indication 
of the localized character of the Ni-C bonds in the nanocrystallites. 

\begin{figure}
\includegraphics[width=85mm]{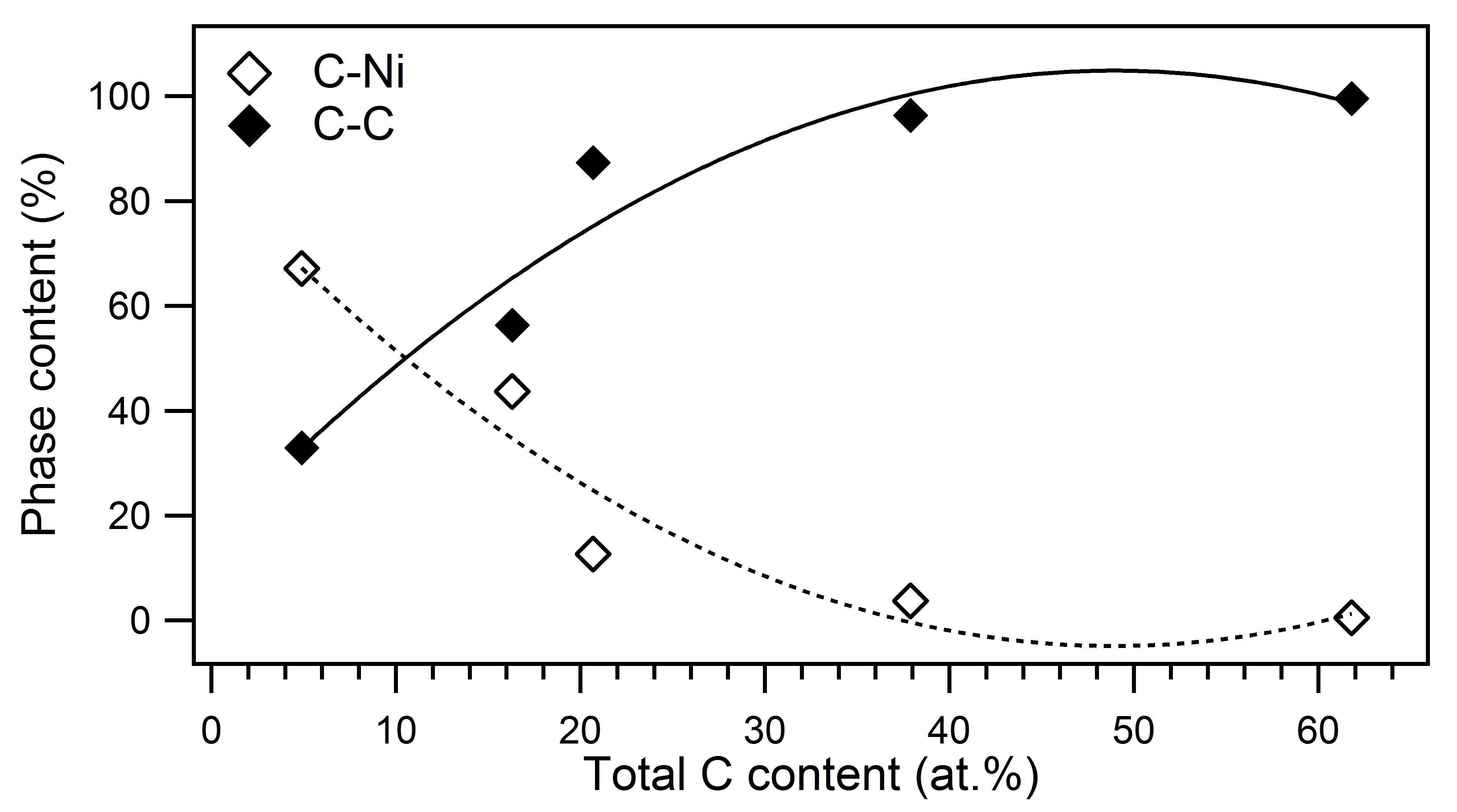} 
\vspace{0.2cm} 
\caption[] {Relative amount of C-Ni, and C-C bonds determined as the proportions 
of the areas fitted on the XPS C \textit{1s} peak. The two curved lines are guidelines 
for the eye.}
\label{XPS2}
\end{figure}

Figure 4 shows the relative amount of the carbide and a-C phases as a function 
of total carbon content (assuming that the Ni-C* contribution to 283.9-284.5 eV 
peak can be neglected). As can be seen, the relative amount of the a-C phase increases 
non-linearly with the total carbon content. The total composition analysis in Table 
I is valid under the assumption that the total photoemission cross section in all 
the samples is constant for carbon. The combined carbon in the fcc-NiC$_{x}$ 
and hcp-NiC$_{y}$ carbide phases can now be estimated using the data 
in Fig. 4 as presented in Table I. The analysis show that the carbon content of 
the carbide phase strongly increase with the total carbon content from 15.7 at\% 
(0.16 at\% total), 36 at\% (0.38 at\% total) to 60 at\% (0.62 at\% total). However, 
the estimated carbon content in the carbide phase represents a lower limit since 
the contribution of Ni-C* has been neglected in the analysis of the C \textit{1s} 
spectra. The variation of the carbon content in the NiC carbide phase is consistent 
with the small dispersion of the Ni $2p_{3/2}$ XPS peak position from 852.7 eV 
(\textit{x}=0.05), 852.9 eV (\textit{x}=0.16), 853.0 eV (\textit{x}=0.38), to 853.1 eV (\textit{x}=0.62).

\begin{figure}
\includegraphics[width=85mm]{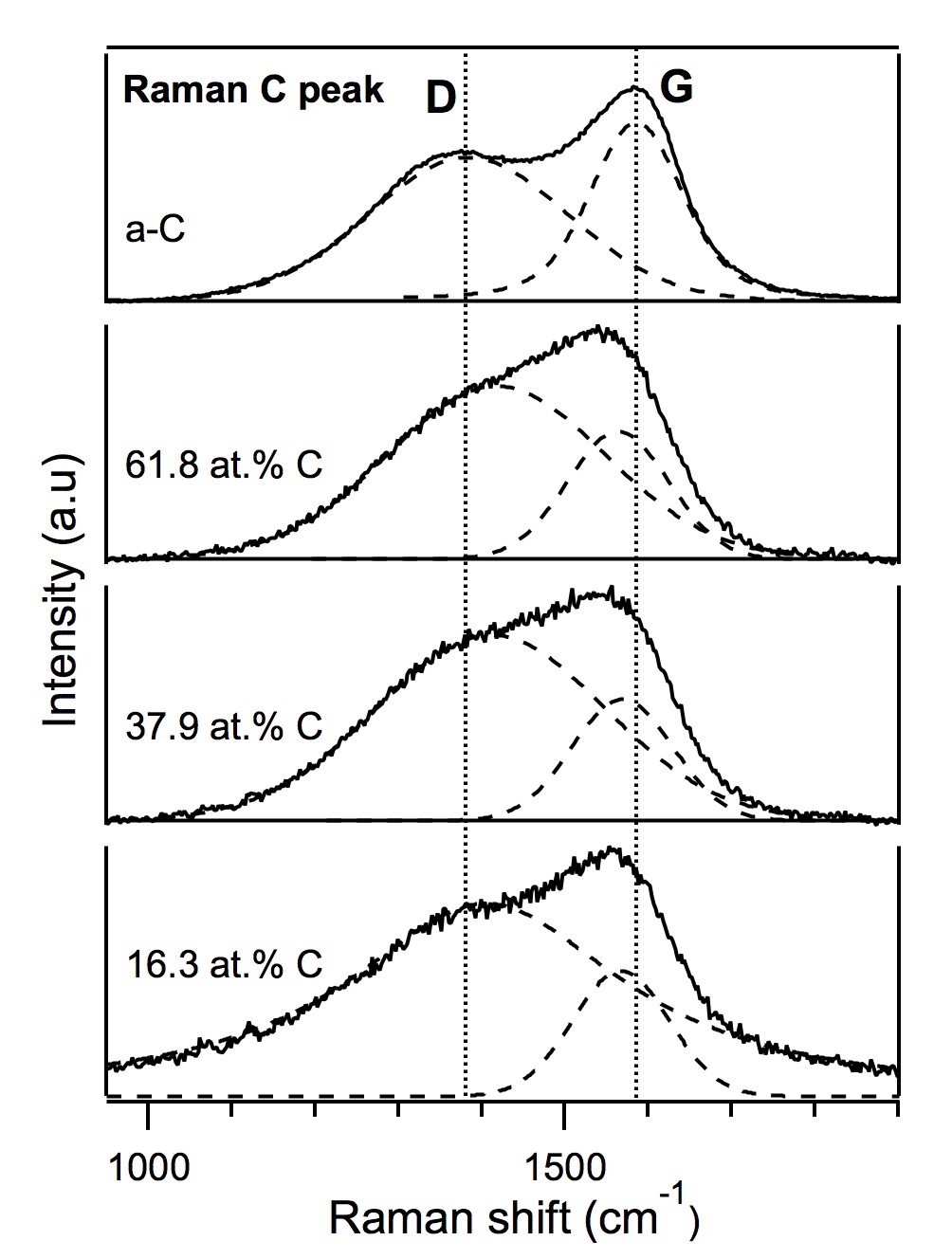} 
\vspace{0.2cm} 
\caption[] {Raman spectra for the carbon peak of the Ni$_{1-x}$C$_{x}$ films for \textit{x}=0.16, 0.38, 0.62, and amorphous carbon (a-C). The two vertical 
dashed lines indicate the disorder (D), and graphite (G) peaks of the fitted peak 
components \cite{27,28}.}
\label{Raman}
\end{figure}

\subsection{Raman spectroscopy}
Figure 5 shows carbon Raman spectra of the Ni$_{1-x}$C$_{x}$ 
films with \textit{x} values of 0.16, 0.38, and 0.62, in comparison to pure amorphous 
carbon (a-C). As the Raman scattering cross section from C-Ni is low, the Raman spectra are dominated by the segregated part of the carbon in the compounds. The two band components in the spectra, the disordered (D), and graphite 
(G) peaks, were deconvoluted by Voigt shape functions. In pure graphite, the vibrational 
mode that gives rise to the G-band is known to be due to the relative motion of 
$sp^{2}$ hybridized C 
atoms while the D band is due to the breathing vibrational mode of the six-membered 
rings \cite{27}. The energies of the D, and the G bands have an almost constant position 
around 1410 cm\textsuperscript{-1}, and 1570 cm\textsuperscript{-1}, respectively. 
Both peaks are slightly shifted together with respect to their positions for the 
pure a-C of 1384 cm\textsuperscript{-1}, and 1585 cm\textsuperscript{-1} respectively.\textsuperscript{ 
}The I$_{D}$/I$_{G}$ height ratios of the films are 1.53, 
1.59, 1.36 while for the a-C film, the ratio is lower (0.87) due to a more graphitic 
character of the carbon bonds \cite{28}. These ratios approximately correspond to $sp^{2}$ 
fractions of 0.68, 0.70, 0.65 while it is lower for a-C (0.59). The predominant 
$sp^{2}$ hybridization is consistent with the observations 
in the XPS spectra of the same samples.

\begin{figure}
\includegraphics[width=88mm]{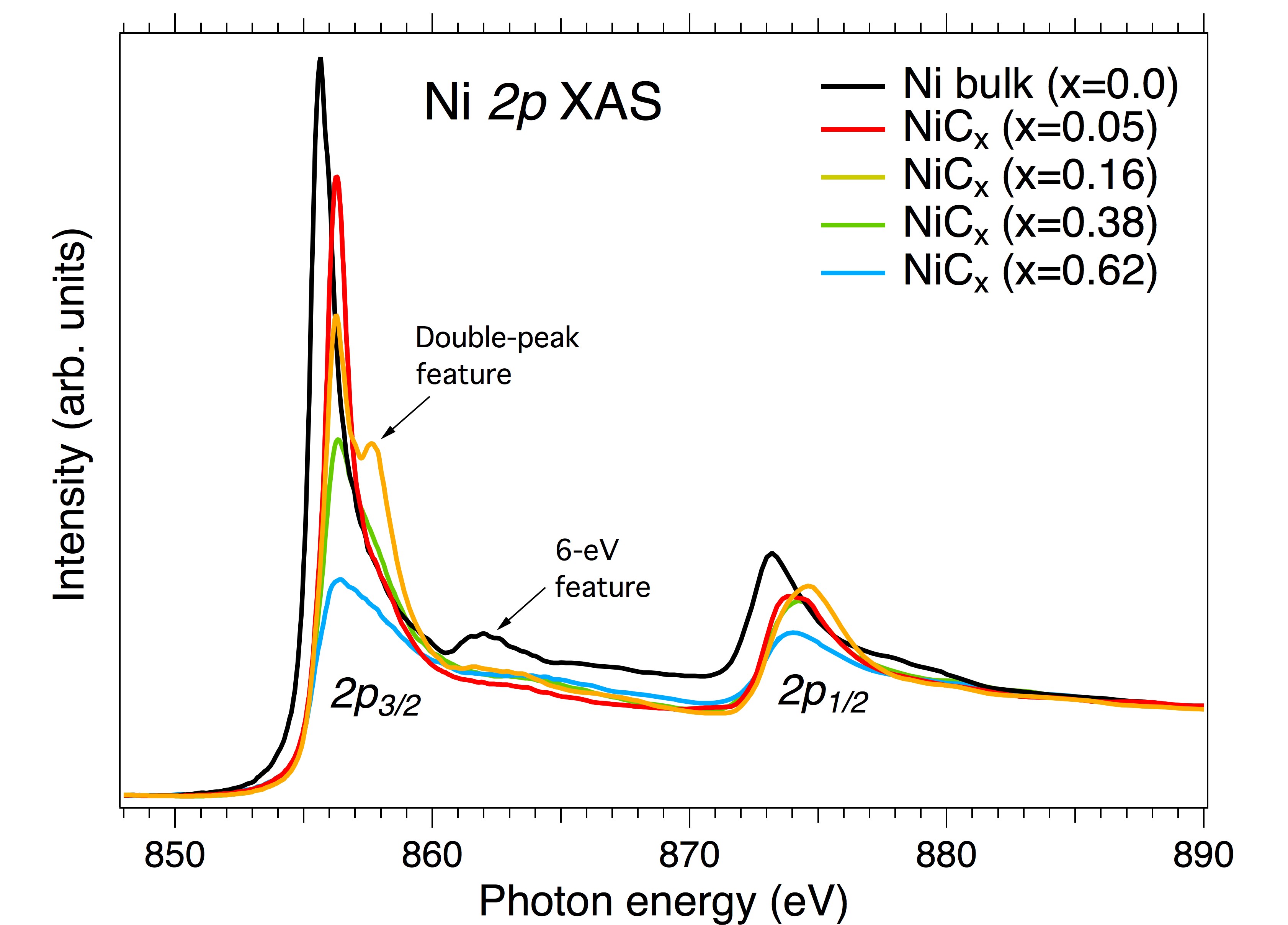} 
\vspace{0.2cm} 
\caption[] {(Color online) Ni $2p$ TFY-XAS spectra of Ni$_{1-x}$C$_{x}$ for different C contents in comparison to bulk Ni (x=0.0).}
\label{Raman}
\end{figure}

\subsection{Ni $2p$ X-ray absorption spectroscopy}
Figure 6 shows Ni $2p$ XAS spectra of the 3\textit{d}, and 4\textit{s} conduction bands following the Ni $2p_{3/2,1/2}$ $\rightarrow$ $3d$ dipole transitions of the Ni$_{1-x}$C$_{x}$ films with different carbon content in comparison to Ni metal. The Ni $2p$ XAS spectra mainly represent the nickel contribution in the fcc-NiC$_{x}$ and hcp-NiC$_{y}$ carbide phases. The main peak structures are associated with the Ni $2p_{3/2}$ and the $2p_{1/2}$ core-shell spin-orbit splitting of 17.3 eV. 
A comparison of the spectra shows four interesting effects: (i) the intensity of the the main $2p_{3/2}$ peak decrease with carbon content. The Ni $2p$ XAS intensity is proportional to the unoccupied $3d$ states, and the intensity trend indicates that the Ni $3d$ electron density decreases around the absorbing Ni atoms for higher carbon concentration. 
The intensity of the normally sharp Ni $2p_{3/2}$ XAS peak in pure crystalline Ni phase, is largely suppressed by the broadening and distribution of different types of chemical bonds.
(ii) For comparison, the XAS spectrum of fcc Ni metal (\textit{x}=0.0) has narrower, and more intense $2p_{3/2}$, and $2p_{1/2}$ absorption peaks, whereas the XAS spectra of the carbon-containing films are broader and shifted by 0.6 eV towards higher photon energy. This energy shift is an indication of higher ionicity of Ni as a result of charge-transfer from Ni to C. (iii) For \textit{x}=0.16, a pronounced double-peak structure with 1.4 eV splitting from the main peak at the $2p_{3/2}$ peak is observed. The double-peak feature is similar to the $t_{2g}$-$e_{g}$ crystal field splitting observed in TiC nanocomposites \cite{16}. It is a signature of a change in orbital occupation to the hcp crystal structure while the sharp single-peak feature of Ni metal is a signature of fcc structure (cubic). For \textit{x}=0.05, the double-peak structure has essentially vanished in comparison to at \textit{x}=0.16 due to the superposition of the strong fcc contribution. (iv) The 6-eV feature \cite{17,29} above the main $2p_{3/2}$ peak is prominent in Ni metal \textit{x}=0.0) that is associated with electron correlation effects and narrow-band phenomena \cite{30}. The intensity of the 6-eV feature is very low in the Ni$_{1-x}$C$_{x}$ films in comparison to Ni metal even at \textit{x}=0.05 due to more delocalized bands.

A comparison of the spectral shapes at different carbon contents shows that the $2p_{3/2}$/$2p_{1/2}$ branching ratio estimated by the peak height is largest (3.0) for the lowest carbon content (\textit{x}=0.05) and is similar as for Ni metal. With increasing carbon content, the branching ratio decreases to 2.3 (\textit{x}=0.16), 1.8 (\textit{x}=0.38) and 1.3 (\textit{x}=0.62). Integration of peak areas by Gaussian functions give the same trend as comparison of the the peak heights but yield a higher branching ratio for Ni metal than for the other samples. A lower $2p_{3/2}$/$2p_{1/2}$ branching ratio is an indication of higher ionicity (lower conductivity) for the highest carbon content \cite{31,32,33}. However, the $2p_{3/2}$/$2p_{1/2}$ branching ratio is a result of the ionicity for Ni, mainly in the fcc-NiC$_{x}$ and hcp-NiC$_{y}$, carbide components, and not for the entire film. For the higher carbon contents, when the main part of the film consists of C-rich matrix areas, this phase determines the resistivity.

\begin{figure}
\includegraphics[width=88mm]{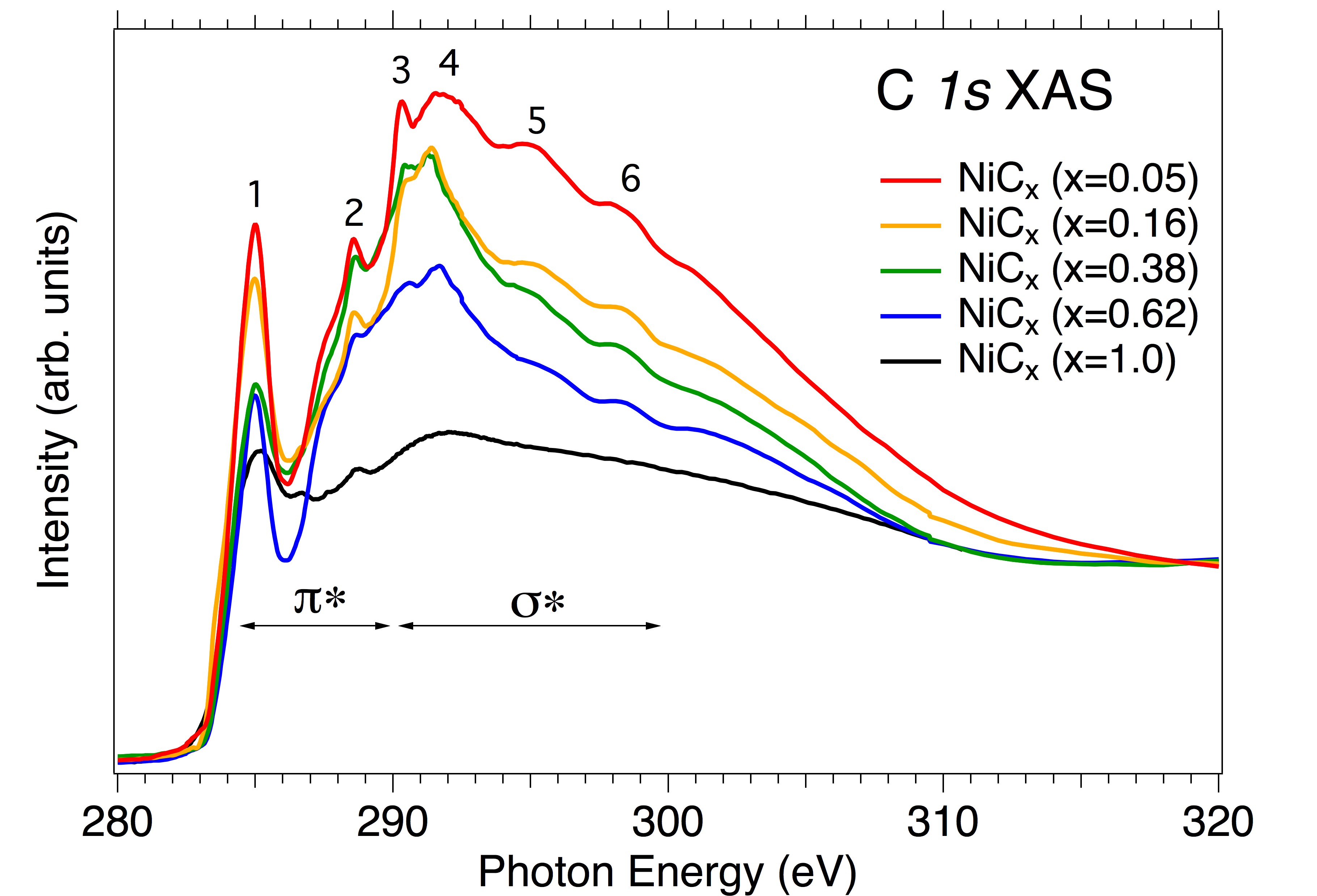} 
\vspace{0.2cm} 
\caption[] {(Color online) C $1s$ TFY-XAS spectra of Ni$_{1-x}$C$_{x}$ 
for different C contents compared to amorphous carbon (a-C, \textit{x}=1.0).}
\label{Raman}
\end{figure}

\subsection{C $1s$ X-ray absorption spectroscopy}
Figure 7 shows C $1s$ XAS spectra of the Ni$_{1-x}$C$_{x}$ films probing the unoccupied C $2p$ conduction bands as a superposition of the NiC carbide phases, and the changes in the C matrix phase with composition. The first peak structure (1) at $\sim$285 eV is associated with empty $\pi$* states, and the higher states (3-6) above 290 eV are associated with unoccupied $\sigma$* states. The empty $\pi$* orbitals (1) consists of the sum of two contributions in Ni$_{1-x}$C$_{x}$: (i) $sp$\textsuperscript{2} (C=C), and $sp$\textsuperscript{1} hybridized C states in the amorphous carbon phase, and (ii) C $2p$ - Ni $3d$ hybridized states in the fcc-NiC$_{x}$ and hcp-NiC$_{y}$ carbide phases. The peak at $\sim$288.5 eV is also due to C \textit{2p} - Ni \textit{3d} hybridization with a superimposed contribution from the carbon phase \cite{16}. The energy region above 290 eV is known to originate from $sp$\textsuperscript{3} hybridized (C-C) $\sigma$* resonances, where the peak (4) at 291.5 eV forms a shape resonance with multielectron excitations towards higher energies \cite{16}. The most intense structure shows highest intensity for $x$=0.05 that originates from $sp^{3}$ hybridized $\sigma$* states. 
Additional $\sigma$* states (5), (6) at 295 eV and 298 eV are also associated with $sp^3$ bonding. Contrary to the case of TiC \cite{16}, there is no pre-peak below the $\pi$* peak at 283.3 eV for NiC. 

The integrated $\pi$\textsuperscript{*}/[$\pi$\textsuperscript{*}+$\sigma$\textsuperscript{*}] intensity ratio was calculated by fitting a step-edge background with a Gaussian function to each peak following the procedure in Refs. \cite{33a,34} in order to not overestimate the  $\sigma$ contribution. We assumed that $\pi$* peaks occur below and $\sigma$* peaks above 290 eV as indicated in Fig. 7. This analysis method gives an estimation of the relative amount of $\pi$\textsuperscript{*}($sp^{2}$, $sp^{1}$ hybridization content in the samples as shown in Table I. The fraction of $sp^{2}$ is smallest for $x$=0.05 and highest for a-C, following a similar trend as the XPS and the Raman results. 

\begin{figure}
\includegraphics[width=85mm]{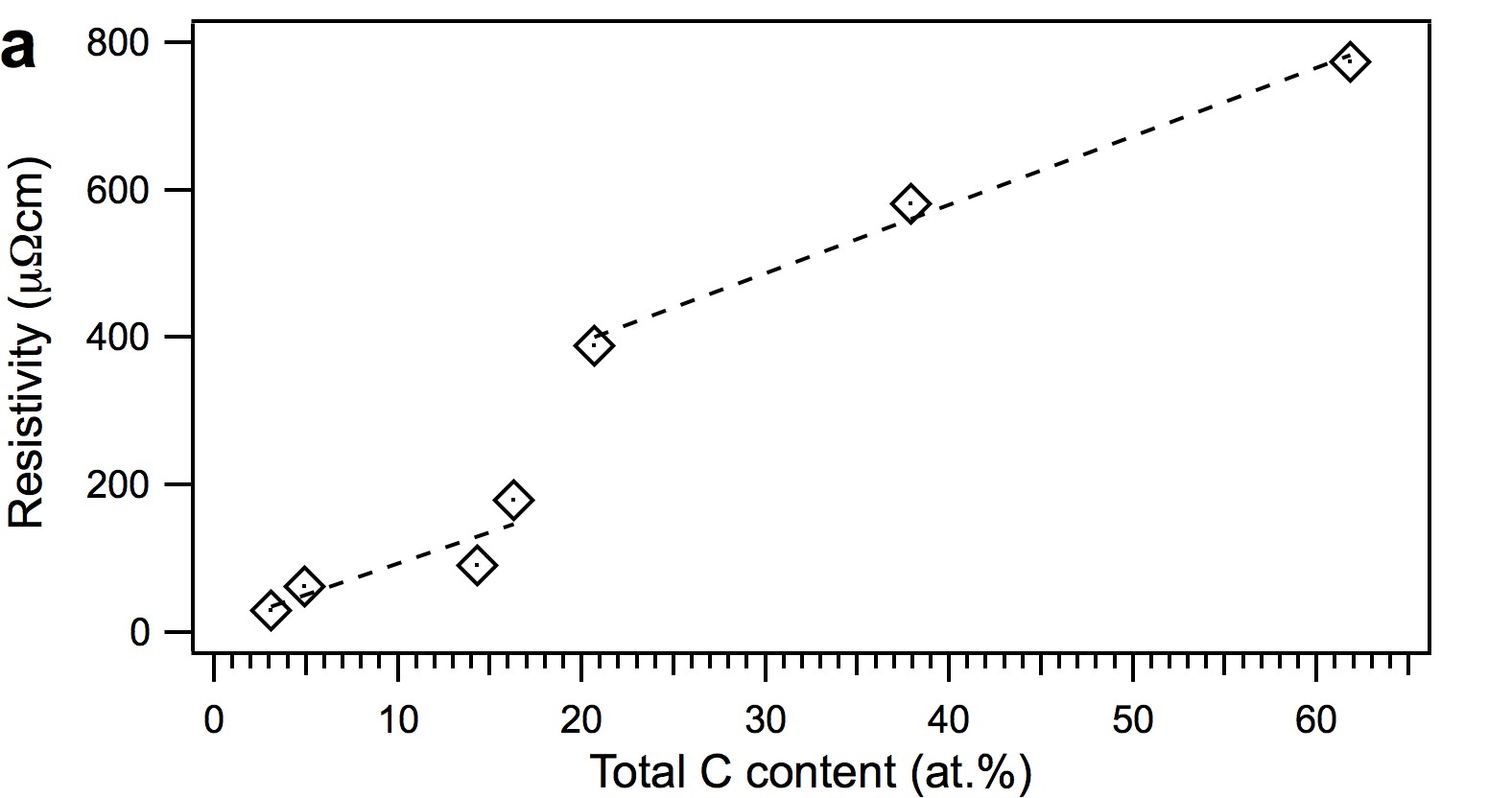} 
\includegraphics[width=85mm]{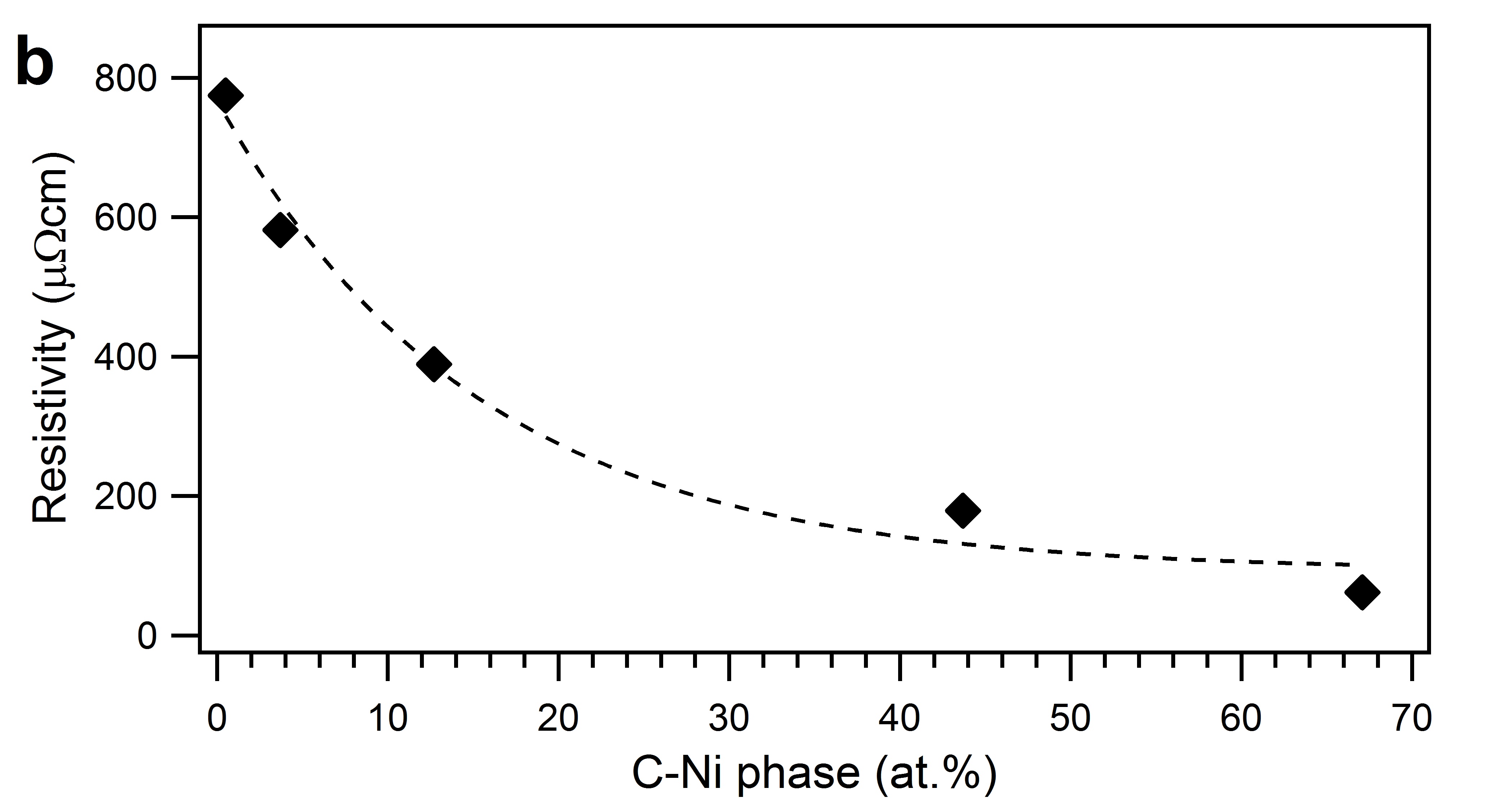} 
\vspace{0.2cm} 
\caption[] {a) Resistivity of the Ni$_{1-x}$C$_{x}$ 
films depending on the C content as determined from sheet resistance measurements 
by a four-point probe and film thickness determined by SEM. b) The resistivity 
is plotted as a function of the relative amount of C bound in the C-Ni bonds. The 
dashed least-square fitted curves are guides for the eye.}
\label{Raman}
\end{figure}

\subsection{Resistivity measurements}
Figure 8a shows the electrical resistivity in the Ni$_{1-x}$C$_{x}$ films as a function of carbon content \textit{x}. Compared to the electrical resistivity 
of 6.93  $\mu\Omega$cm \cite{35} for Ni metal, the small introduction of C of $\sim$3 
at.\% increases the resistivity to $\sim$30 $\mu\Omega$cm. However, it is 
well known that metallic thin films display higher electrical resistivity compared 
to the bulk metals \cite{35,36}. When increasing the C content another 2 at.\% C, the 
resistivity doubles to $\sim$62 $\mu\Omega$cm, and continues approximately 
linearly to $\sim$150 µΩcm at 16.3 at\% C. Above this carbon content, 
the samples transform from polycrystalline to amorphous and the resistivity increases 
by more a factor of two to $\sim$400 $\mu\Omega$cm. Above 20.7 at\% C, the 
resistivity increases approximately linearly up to a maximum value of $\sim$775 
$\mu\Omega$cm for the C content of 61.8 at.\%. The increase of electrical resistivity 
with increasing C content in the films is correlated to the increased amount of 
C-C bonding (Fig. 4), and the interstitial incorporation of inter-bonded C atoms 
between the Ni lattice sites forming the fcc-NiC$_{x}$ and hcp-NiC$_{y}$ 
carbide phases. The a-C phase is known to be a very poor conductor compared to 
Ni metal. As the XPS analysis shows contributions from C-C bonds for all the investigated 
films, an increase of resistivity is not only governed by the increase of the amorphous 
C phase but is also influenced by the crystallinity. Fig. 8b shows the exponential 
decrease of electrical resistivity with the increasing proportion of the NiC carbide 
phases. The symmetry of the C-Ni component of the C \textit{1s} XPS peak also suggests 
a low electrical conductivity of the carbide. However, as shown by the XPS analysis, 
there is only a small amount of C incorporated into Ni. Therefore, the NiC phases 
could be regarded as a solid solution \cite{36,37} with low amount of C solved into 
Ni rather than as an ordinary carbide phase. Thus, the C-Ni component of the structure 
is mostly metallic, leaving the a-C matrix phase to determine the general trend 
in the electrical resistivity of the films.   

\begin{table*}[tp][h]
\caption[tabbetas]{\label{tab:sml.1}\sf Composition of the Ni-C films for $x$=0.05, 0.16, 0.38, and 0.62. 
The amount of carbon in the carbide phase and the \textit{sp}\textsuperscript{\textit{2}} 
fractions were determined by integrating the areas under the corresponding peak 
structures in C \textit{1s} XPS spectra. The \textit{sp}\textsuperscript{\textit{2}} 
fractions in Raman were estimated from Ref. 28.}
\[\begin{array} {l c c c c c}
\hline
\multicolumn{1}{c}{\mbox{Total composition}}&\multicolumn{1}{c}{\mbox{Ni$_{0.95}$C$_{0.05}$}}&\multicolumn{1}{c}{\mbox{Ni$_{0.84}$C$_{0.16}$}}&\multicolumn{1}{c}{\mbox{Ni$_{0.62}$C$_{0.38}$}}&\multicolumn{1}{c}{\mbox{Ni$_{0.38}$C$_{0.62}$}}&\multicolumn{1}{c}{\mbox{a-C}}\\
\hline
{\mbox{at\% C in NiC$_y$ phase}}&5.0&15.7&36&60&100\\
{\mbox{XPS $sp^2$ fraction}}&0.64&0.80&0.69&0.70&-\\
{\mbox{Raman $sp^2$ fraction}}&-&0.77&0.79&0.89&0.71\\
{\mbox{C $1s$ XAS $\pi$*/[$\pi$*+$\sigma$*]}}&0.42&0.53&0.60&0.56&0.72\\
\hline
\end{array}\]
\end{table*}

\section{Discussion}
The most significant difference between Ni$_{1-x}$C$_{x}$ 
as compared to other late transition-metal carbides such as Cr$_{1-x}$C$_{x}$ 
\cite{4}, and Fe$_{1-x}$C$_{x}$ \cite{38}, is the precipitation of 
Ni-based phases in the form of nanocrystals, while Cr, and Fe form mainly completely 
amorphous films for a large range of compositions. In our combined analysis of 
XRD, and HR-TEM, we find that at low carbon content (\textit{x}=0.05), the Ni$_{1-x}$C$_{x}$ 
films consist of hcp-NiC$_{y}$ nanocrystals with a smaller contribution 
(25-30\%) of fcc-NiC$_{y}$ nanocrystals, where \textit{y}$\leq$0.30. For 
low carbon contents (\textit{x}=0.05 and 0.16), the estimated average grain size 
is relatively large, 10-20 nm. 

Previous investigations of hcp-Ni nanocrystals are lacking a material composition 
analysis \cite{6,7}, but pure hcp-Ni metal is known to be unstable or metastable. To 
our knowledge, only one previous experiment showed that crystallites of hcp-Ni 
metal could be synthesized, and it was easily transformed into fcc-Ni metal when 
its size was larger than 5 nm \cite{11}. Most of the reported hcp-Ni metals are likely 
stabilized by carbon. The Ni atom in rhombohedral-Ni$_{3}$C and hcp-Ni 
structures occupy exactly the same positions and yield identical XRD data at high 
angle (\texttt{>}38\textsuperscript{o}) and, this is the reason why it has not 
been identified before. Recently, Schaefer \textit{et al.} pointed out that it 
is possible to experimentally distinguish hcp-Ni and rhombohedral-Ni$_{3}$C 
structures by using low-angle XRD. In addition, He and Schaefer claimed that there 
exists no hcp-Ni because they inferred that all the previously reported hexagonal 
Ni carbides contained carbon, presented 01-12 and 1-10-4 reflections, and should 
therefore be described as rhombohedral Ni$_{3}$C \cite{13,14}. However, in 
our sputter-deposited Ni$_{1-x}$C$_{x}$ samples, the absence 
of 01-12, and 1-10-4 reflections indicates that the superstructure of rhombohedral 
Ni$_{3}$C is not formed, and instead, a hcp-Ni structure occurs. It 
should be noted that our hcp-Ni structure does contain C and its cell parameters 
depend on the carbon content. This is consistent with previous works, showing that 
hcp-Ni is stabilized by carbon. Thus, our hcp-Ni phase should be described as hcp-NiC$_{y}$ 
instead of hcp-Ni or rhombohedral-Ni$_{3}$C$_{1-x}$. 
Uhlig \textit{et al.} also found a carbon containing Ni structure \cite{9}.
However, the absence of low-angle reflections indicates that their films consists of hcp-Ni with a carbon content rather than Ni$_{3}$C.

The XPS and XAS measurements confirm the structure to be carbidic and do not show 
spectral profiles of metallic Ni. From a structural point of view, the difference 
between hcp-Ni$_{3}$C and rhombohedral Ni$_{3}$C is the carbon 
position: ordered interstitial C in rhombohedral Ni$_{3}$C and disordered 
interstitial C in hcp-Ni$_{3}$C. The formation of hcp-NiC$_{y}$ 
instead of rhombohedral-Ni$_{3}$C may be due to the non-equilibrium 
sputtering process. Moreover, at low carbon content, hcp-NiC$_{y}$ or 
rhombohedral-Ni$_{3}$C is likely more stable than fcc-NiC$_{x}$. 
Further theoretical calculations will be performed to verify this hypothesis, and 
consequently give an interpretation why NiC$_{x}$ form crystalline phases, 
whereas CrC$_{x}$ and FeC$_{x}$ form amorphous phases at 
low carbon content. 

The difference in XPS binding energy of the C-Ni peak in comparison to the C-C peak (2.0 eV) is due to the different types of bonding environments. A small low-energy shift of 0.15 eV between the samples with 4.9 at
This observation is consistent with a small XPS high-energy shift of 0.25 eV at the Ni $2p_{3/2}$ edge for these crystalline samples. As the structure of the samples change from crystalline to amorphous at 38 and 62 at
Fe, one would also expect a smaller chemical shift in the case of Ni carbides. 
This scenario with a smaller chemical shift for the late transition metal carbides 
is consistent when comparing to Ti-C (2.5 eV) \cite{5} but not for Fe-C (1.7 eV) \cite{38} 
and Cr-C (1.5 eV) \cite{3}, where it is smaller. In this respect, XAS gives important 
complementary information to XPS about charge-transfer effects. The general intensity 
trend in the $2p$ branching ratio of the Ni XAS spectra is a signature of charge-transfer from Ni to C that 
is largest for the sample that contains most C (i.e., \textit{x}=0.62). The 
variation in intensity of the unoccupied states reflects changes in orbital occupation 
and bonding of the atoms at the carbide/matrix interface between crystallites, 
and amorphous domains. It can be assumed that charge-transfer occurs within the 
fcc-NiC$_{x}$ and hcp-NiC$_{y}$ nanocrystal carbide phases, 
but more significant across the carbide/matrix interface with the surrounding amorphous 
C-phase or between nanocrystals, that depends on the nanocrystalline size. 

Moreover, the 6-eV feature in the Ni XAS spectra that signifies electron correlation 
effects and narrow-band phenomena in metallic Ni \cite{17,29} is washed out in the Ni$_{1-x}$C$_{x}$ 
samples due to the Ni $3d$-C $2p$ orbital overlap that changes the 
properties of Ni already at very low carbon content. Thus, the spectral profiles of the Ni$_{1-x}$C$_{x}$ samples 
exhibit carbide signatures and exclude metallic nickel. Furthermore, for the carbon 
content of \textit{x}=0.16, the ligand-field type of splitting by 1.4 eV that occurs 
in Ni XAS signifies a change in the local coordination and orbital occupation with the formation of 
the single-phase hcp-NiC$_{y}$ carbide phase. The most stable NiC phase 
is cubic \cite{15}, but from the combined shape of the Ni $2p$ (no crystal-field splitting) and C $1s$ XAS spectra (absence of pre-peak), 
this is excluded and consistent with the XRD observations. Using surface-sensitive TEY measurements, Choo \textit{et al.} \cite{39} associated the double-structure in Ni 2p XAS of hcp Ni with surface oxidation. With bulk-sensitive TFY-XAS, we find that this feature is due to the intrinsic hcp Ni structure.

Since charge-transfer effects are clearly observed in bulk-sensitive XAS, this 
contribution is also expected in the more surface sensitive XPS spectra. Although 
this contribution is difficult to separate in the XPS data, part of the third peak 
between the C-C, and C-Fe peaks should be associated with charge-transfer effects at the interfaces between nanocrystals. 
The size and the number of fcc-NiC$_{x}$ and hcp-NiC$_{y}$ 
nanocrystals affect the amount of interface, and charge-transfer between the domains. 
As observed by the XPS analysis, we find that the carbon content in the carbide 
phase varies significantly with the total carbon content (Table I). However, the 
amount of $sp^{2}$-fraction (0.6-0.8) as observed 
in XPS, Raman and XAS is higher in all samples in comparison to a-C and does not 
change significantly when the samples transform from crystalline to amorphous. 
On the other hand, the trend in the resistivity depends on the carbon content as 
well as the crystallinity. To energetically explain why particular nanocrystals 
form in the Ni-C system and not in other late transition metal carbides, further 
experimental and theoretical work will be carried out including the effect of magnetic 
properties.

\section{Conclusions}
Magnetron sputtered nanocomposite Ni$_{1-x}$C$_{x}$ films 
were investigated for a large composition range (0.05$\leq$x$\leq$0.62). We discovered 
a novel hcp-NiC$_{y}$ phase, and show how it is different from rhombohedral 
Ni$_{3}$C. At low carbon content (4.9 at\%), the Ni$_{1-x}$C$_{x}$ 
film consists of hcp-NiC$_{y}$ and fcc-NiC$_{x}$ nanoparticles 
with an average grain size of 10-20 nm as observed by high-resolution X-ray diffraction 
and transmission electron microscopy. With increasing carbon content (16 at\%), 
single-phase hcp-NiC$_{y}$ is formed also with an average grain size 
of 10-20 nm. A double structure in the X-ray absorption spectra reveals a change 
in the orbital occupation and bonding for hcp-NiC$_{y}$ in comparison 
to the other samples. Further increase of carbon content to 38 at\%, and 62 at\%, 
transforms the films to complex X-ray amorphous materials with a mixture of randomly-oriented 
short-range ordered hcp-NiC$_{y}$ and fcc-NiC$_{x}$ nanodomain 
structures surrounded by an increasing amount of amorphous carbon-rich matrix. 
X-ray photoelectron and X-ray absorption spectroscopy analyses reveal that interbonding 
states between the nanocrystallites, and domain structures represent 
a third type of phase that increases with carbon content. The general trend of 
increased electrical resistivity with increasing carbon content in the films is 
correlated to the increased amount of C-C bonding observed in X-ray photoelectron 
spectroscopy. As the C-Ni phase component of the structure is metallic, the carbon 
matrix phase determines the electrical resistivity of the films when the films 
are amorphous. We also find an increase of the resistivity by more than a factor 
of two when the samples transform from crystalline to amorphous.\\

\section{Acknowledgements}
We would like to thank the staff at the MAX IV Laboratory for experimental support, 
and Jill Sundberg, UU, for help with the Raman measurements. The work was supported 
by the Swedish Research Council (VR) Linnaeus, and Project Grants. M. M., U. J. 
and J. L. also acknowledges support from the SSF synergy grant \textit{FUNCASE} Functional 
Carbides and Advanced Surface Engineering.\\

$^{**}$Corresponding author: Martin.Magnuson@ifm.liu.se

\end{document}